%% file: arxiv.tex
\documentclass[sigplan,twocolumn,nonacm]{acmart}

\renewcommand\footnotetextcopyrightpermission[1]{}

\AtBeginDocument{%
  }

\usepackage{microtype}
\usepackage{graphicx}
\usepackage{comment}
\usepackage{soul, color, colortbl}
\usepackage{graphicx}
\usepackage{caption}
\usepackage{subcaption}
\usepackage{threeparttable}
\usepackage{multirow}
\usepackage{booktabs}
\usepackage{verbatim}
\usepackage{epstopdf}
\usepackage{rotating}
\usepackage{listings}
\usepackage{listing}
\usepackage{paralist}
\usepackage{tabularx}
\usepackage{tabu}
\usepackage{arydshln}

\usepackage{amssymb}
\usepackage[shortlabels]{enumitem}
\usepackage[ruled,vlined,linesnumbered]{algorithm2e}
\usepackage{balance}
\usepackage{endnotes}
\usepackage{xspace}
\usepackage{amsfonts}
\usepackage{pifont}
\usepackage{wasysym}
\usepackage{tikz}
\usepackage{xcolor}
\usepackage{float}
\usepackage{svg}
\usepackage{multirow}

\usepackage{hyperref}
\definecolor{darkblue}{rgb}{0.0, 0.0, 0.55}
\definecolor{darkcandyapplered}{rgb}{0.64, 0.0, 0.0}
\definecolor{mygray}{gray}{0.9}
\definecolor{dkgreen}{rgb}{0,0.6,0}
\definecolor{mauve}{rgb}{0.58,0,0.82}

\hypersetup{
    colorlinks=true,
    linkcolor=darkblue,
    citecolor=darkcandyapplered,
    filecolor=magenta,      
    urlcolor=darkblue,
}

\newcommand{\sys}{\textsl{xTier}\xspace}

\begin{document}


\title{\sys: Intelligent Tiering for CXL-Enabled Memory}

\setlength{\textfloatsep}{8pt plus 2pt minus 2pt}
\setlength{\floatsep}{6pt plus 2pt minus 2pt}

\author{Sriranga Ramaswamy}
\affiliation{%
  \institution{University of Colorado Boulder}
  \city{Boulder}
  \state{Colorado}
  \country{USA}
}
\email{sriranga@colorado.edu}

\author{Yueqi Chen}
\affiliation{%
  \institution{University of Colorado Boulder}
  \city{Boulder}
  \state{Colorado}
  \country{USA}
}
\email{yueqi.chen@colorado.edu}

\input{sections/abstract}
\maketitle

\makeatletter
\fancyhead{}%
\fancyhead[LO,LE]{\ACM@linecountL}%
\fancyhead[RO,RE]{\ACM@linecountR}%
\renewcommand{\headrulewidth}{0pt}%
\makeatother

\input{sections/intro}
\input{sections/background}
\input{sections/motivation}
\input{sections/design}
\input{sections/impl}
\input{sections/eval}
\input{sections/discussion}

\input{sections/related}
\input{sections/conclusion}

\newpage
{
\bibliographystyle{ACM-Reference-Format}
\bibliography{reference.bib}{}
}
\newpage
\appendix
\input{sections/appendix}

\end{document}

%% file: sections/abstract.tex
\begin{abstract}

CXL-enabled memory expands server memory capacity, but introduces a page-placement problem: the operating system must decide which pages should reside in DRAM and which should reside on slower CXL memory. Existing systems make this tradeoff in one of two ways. Userspace controllers support flexible policies, but expose placement decisions to scheduler jitter and kernel-userspace crossing overhead. Kernel-space systems avoid this latency, but rely on fixed heuristics that must generalize across workloads.

We present \sys, a kernel-resident learned memory-tiering system. \sys attaches eBPF programs to PEBS events and uses a compact quantized MLP to score sampled pages inside the kernel at microsecond-scale latency. Rather than reacting to every candidate, \sys converges to a low-churn placement for the current workload phase, reduces sampling cost after convergence, and returns to a higher sampling cadence when the workload shifts.

We evaluate \sys on six memory-bound workloads at DRAM:CXL ratios from 1:5 to 1:25. The advantage grows as the DRAM budget tightens. At 1:15 and beyond, \sys is the fastest system in 14 of 18 configurations. Where it is not fastest, it trails the best baseline by 3.9\% on average. It reaches this performance while moving 13$\times$ fewer pages in geometric mean, and 22$\times$ fewer at the tighter ratios. When a workload changes phase, \sys rebuilds its hot set in DRAM faster and more completely than any baseline. 


\end{abstract}

%% file: sections/intro.tex
\section{Introduction}

Compute Express Link (CXL) memory expansion is reaching production deployment at hyperscale operators~\cite{TPP, IDT,dcxl,artmem}. 
A directly attached CXL memory device appears to the operating system as a CPU-less NUMA node, with load latency roughly 1.5--2$\times$ that of local DRAM~\cite{Pond, FreqTier}. 
CXL therefore allows a server to scale memory capacity beyond its local DIMM slots while preserving cache-line-granularity load/store semantics~\cite{TPP}. 
This capacity comes with a placement problem.
As a workload runs, the operating system must continuously decide which pages deserve scarce DRAM capacity and which can remain on slower CXL memory~\cite{Memtis, FreqTier}.

\begin{figure}[t]
  \centering
  \includegraphics[width=.9\columnwidth]{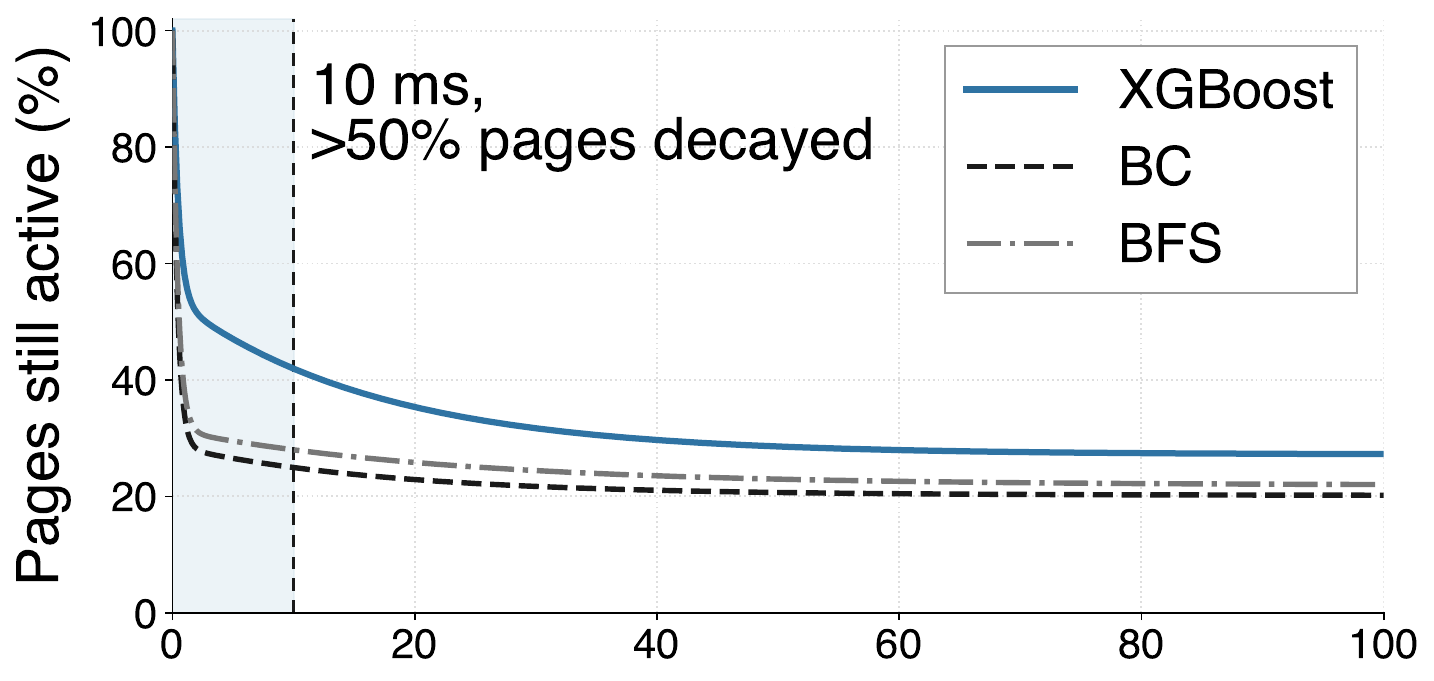}
  \caption{Page hot-residency decay across three workloads since the last access (ms).}
  \label{fig:decay}
\end{figure}

Effective placement must satisfy two requirements simultaneously. 
The first is timeliness. 
Modern memory-intensive workloads can change their access patterns on millisecond timescales, so a page that is hot now may no longer justify migration only a few milliseconds later~\cite{HeMem, Memtis}. 
A placement decision that arrives after the hot set has shifted is therefore stale by construction. 
Figure~\ref{fig:decay} quantifies how quickly this opportunity window closes.
The second requirement is bounded overhead. Migrating a page incurs tens of microseconds of unmapping, copying, and TLB-shootdown cost~\cite{Nimble, TPP}.
At sufficient volume, migration overhead can erase the benefit of placing data in the faster tier~\cite{Memtis}.
The decision-making path itself also consumes CPU cycles that would otherwise be available to the workload.

\begin{figure}[t] \centering       
    \includegraphics[width=.9\columnwidth]{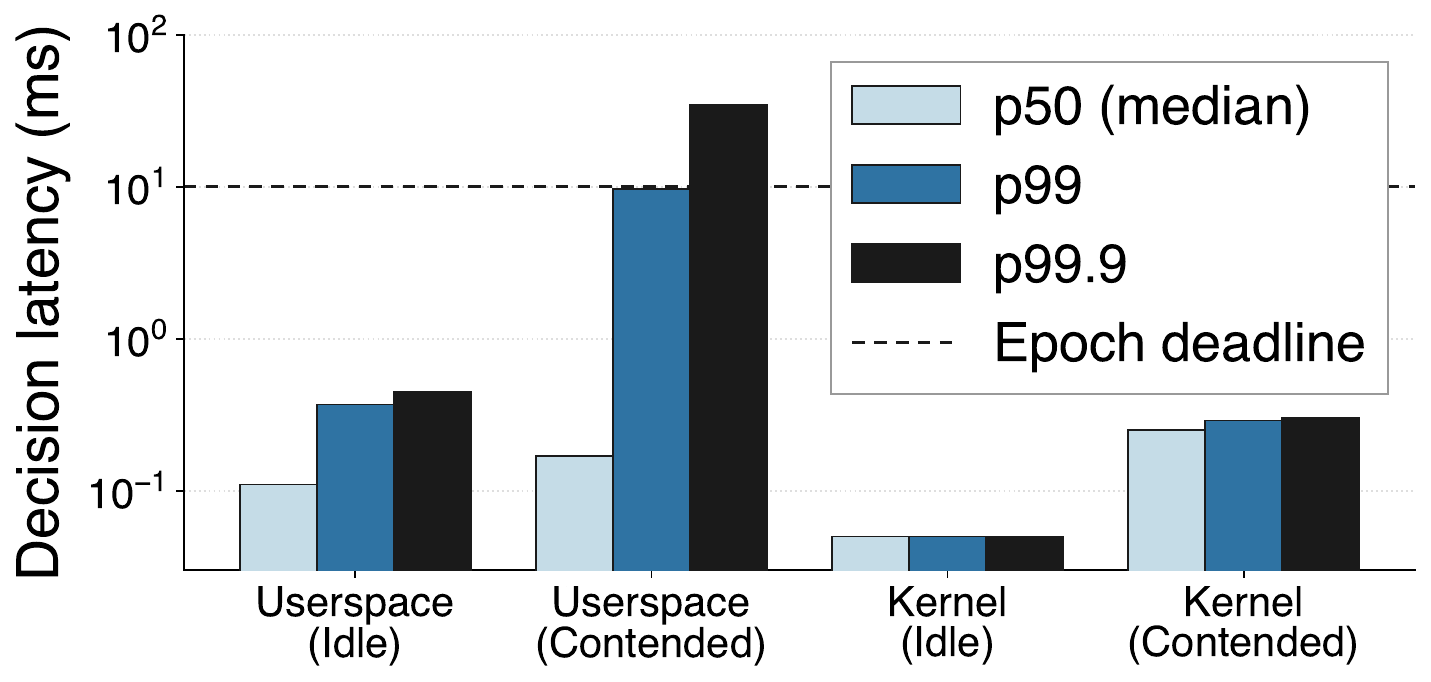} 
    \caption{Per-epoch decision latency under idle and contended conditions. The y-axis is logarithmic.} 
    \label{fig:us_vs_ks} 
\end{figure}

Prior tiered-memory systems approach this problem from two directions. Userspace controllers such as HeMem~\cite{HeMem} sample memory accesses through PEBS and make placement decisions in a userspace daemon. This approach supports arbitrary policy logic, including learned models, but the control loop is exposed to scheduler delay. Figure~\ref{fig:us_vs_ks} shows a userspace path missing the deadline under contention while an in-kernel path stays well inside it. The gap remains when we strip the policy down to an 
\(O(1)\) operation, so it is the boundary crossing that costs, not the policy. 

Kernel-resident systems such as TPP~\cite{TPP}, Memtis~\cite{Memtis}, AutoNUMA~\cite{AutoNUMA}, and TMTS~\cite{TMTS} execute placement logic in the kernel. They reduce control-loop latency, but typically expose a fixed policy whose thresholds or ranking rules must work across diverse workloads~\cite{IDT}. Neither camp provides a placement path that is both kernel-fast and workload-adaptive. Recurring datacenter workloads, which dominate the CXL deployment context~\cite{IDT}, can in principle support per-workload policy calibration. The same binaries run repeatedly on the same class of inputs, so a one-time profiling and training cost per workload amortizes across every later run. No existing system, however, is positioned to execute such a policy at the latency required.

The design point we target follows directly from this gap. Placement decisions must be made inside the kernel, on the same code path as the hardware events that produce them. The ranking function that drives those decisions must be learned from the workload rather than fixed at design time. The cost of migration must be bounded by the number of pages whose movement is actually worth the migration overhead, not by the rate at which candidates are produced.

This work presents \sys, a tiered-memory controller that attaches an eBPF program to PEBS events and evaluates placement decisions in the kernel as hardware samples arrive. Cheap filters discard most samples before any model runs. A quantized multilayer perceptron then ranks the surviving candidates at microsecond scale. An offline training pipeline produces a per-workload model that the loader installs into BPF maps at startup. Deploying a retrained model requires no change to the eBPF object and no kernel rebuild. The architecture also separates promotion and demotion paths, using PEBS to drive promotion decisions and the page-table Accessed bit to drive demotion, so that the cost of each path matches the value of the decisions it produces.

Realizing this design requires working around three constraints.
  The eBPF verifier imposes a per-program instruction limit that a complete MLP forward pass exceeds, so the inference path is split across tail-called programs.
  eBPF also lacks floating-point arithmetic, so weights and activations are quantized to 8-bit integers and the forward pass uses only integer multiply-accumulates.
  Finally, migration must be throttled, or the system moves more traffic than the workload itself. \sys throttles migration with per-page cooldowns, per-epoch admission gates, and a sampling controller that slows once placement converges.

We evaluate \sys against AutoNUMA, TPP, Memtis, and FreqTier on six memory-bound workloads across DRAM:CXL ratios from 1:5 to 1:25. 
The advantage grows as the DRAM budget tightens. 
At 1:15 and beyond, \sys is the fastest system in 14 of 18 configurations, and it trails the best baseline by 3.9\% on average where it is not. 
It reaches this performance while moving 13$\times$ fewer pages in geometric mean, and 22$\times$ fewer at the tighter ratios. 
Under phase changes, \sys rebuilds the DRAM hot set faster and more completely than any baseline. 
\sys consumes under 4\% of one core in steady state.
The key contributions of this paper are:
\begin{itemize}
    \item We present \sys, a kernel-resident learned memory-tiering system that combines workload-specific page ranking with low-latency in-kernel placement decisions for CXL memory.
    \item We design an eBPF-compatible inference path that executes a compact neural ranking model in the kernel using INT8 quantization and verifier-safe partitioning across tail-called programs.
    \item We identify and address a key mismatch in learned tiering. A high candidate rate must not translate into a high migration rate. \sys separates ranking from admission and combines top-$K$ selection, per-page cooldown, Accessed-bit demotion, and adaptive sampling to bound migration traffic while preserving responsiveness.
\end{itemize}

%% file: sections/background.tex
\section{Background}

\subsection{Access tracking for page placement}
Commodity x86 systems provide two complementary page-access signals.
\textbf{PEBS} (Precise Event-Based Sampling)~\cite{pebs} periodically
records sampled memory events, including the accessed virtual address,
data source, and load latency. 
It provides precise positive evidence
of page activity, but its interrupt and record-processing overhead
increases with the sampling rate.

The page-table \textbf{Accessed bit} is set by hardware when a page is
referenced and remains set until cleared by software. 
Page-table scans
can therefore determine whether a page has been accessed since its
preceding observation, but not how often or how recently. 
This signal
is less precise than PEBS but can be collected incrementally at a cost
that does not scale with the number of memory references.

PEBS is consequently well suited to identifying active CXL pages for
promotion, whereas Accessed bits that remain clear across successive scans provide inexpensive
evidence that a DRAM page is a candidate for demotion.

\subsection{eBPF for in-kernel policy}
Extended BPF (eBPF)~\cite{eBPF} enables user-supplied programs to run at kernel event hooks, including hardware performance-counter events.
  Before loading a program, the kernel verifier checks memory safety, bounded control flow, and stack use, and accepted programs are JIT-compiled to native code.
  Persistent state and communication with userspace use typed BPF maps, such as arrays, hash tables, per-CPU storage, and ring buffers.
  The verifier also bounds program size, stack space, and control-flow complexity, and provides no native floating-point arithmetic.
  These limits suit bounded policies close to kernel event sources, but make conventional machine-learning inference difficult.
  Section~\ref{sec:impl} describes how \sys works within them.

\subsection{CXL memory tiering}
Compute Express Link (CXL)~\cite{CXL} exposes byte-addressable memory over PCIe-class links as an additional NUMA node, so ordinary loads and stores reach it transparently. The latency gap is not uniform. Production CXL 1.1 systems report roughly 2$\times$ DRAM latency at idle and 2.9$\times$ under bandwidth pressure~\cite{Pond,FreqTier}, so the cost of misplacement rises with load. Tiering systems therefore keep frequently accessed pages in DRAM and use CXL for the remaining capacity, promoting and demoting pages as access patterns change.

%% file: sections/motivation.tex
\section{Motivation}
\label{sec:motivation}
A tiering system is not just identifying frequently accessed pages. 
This section uses workload measurements and control-path experiments to illustrate the timing, execution, and policy requirements for effective memory tiering.

\subsection{Tiering must be both timely and selective} 
Scanner-based tiering policies typically sweep page tables every hundreds of milliseconds or seconds to bound monitoring and migration costs.
Figure~\ref{fig:decay}, however, shows that page activity evolves much faster.
Median hot residency is on the order of single-digit milliseconds, and most bursts expire within a few tens of milliseconds.
At a 100\,ms cadence, a policy may act only after the activity that justified migration has ended. 
Reacting quickly alone is insufficient because migration itself incurs page isolation, allocation, copying, page-table updates, and TLB invalidation. A short burst may therefore end before the migration cost is recovered.
Promoting every recently active page can therefore cost more than the resulting placement saves.

\vspace{4pt}
\noindent \textbf{Takeaways.} 
These observations establish two requirements for effective tiering. 
First, placement decisions should complete within a 10\,ms window, which captures many observed bursts while leaving time to batch placement work. 
Second, timeliness must be paired with selectivity.
A policy should migrate only pages whose expected future accesses are sufficient to amortize the cost of migration.

\subsection{Latency-critical operations must be in the kernel}

Userspace offers substantial flexibility for implementing placement policies and has therefore been widely adopted by prior systems~\cite{kleio, HeMem, pageflex}. 
However, we find that a userspace control path cannot reliably satisfy the decision window identified above.

Figure~\ref{fig:us_vs_ks} compares the per-epoch latency of userspace and in-kernel placement paths against the 10\,ms target.
On an idle system, both paths meet the target. Under CPU contention, however, the p99.9 latency of the userspace path increases to 34.7\,ms, whereas the in-kernel path remains below the target at 0.30\,ms. 
This gap persists even when the policy computation is replaced with an \(O(1)\) operation, showing that policy evaluation is not the dominant source of delay. 
Instead, a userspace controller must transfer observations out of the kernel, wait for its control thread to be scheduled, and submit placement decisions back to the kernel. 
Scheduler delays, context switches, and event serialization can therefore delay a decision even when the policy itself is inexpensive.

\vspace{4pt}
\noindent \textbf{Takeaways.} 
These results motivate a split architecture. Placement decisions run in the kernel, close to the events that trigger them. Profiling and coarse coordination stay in userspace, where flexibility matters more than latency.

\subsection{Page ranking should be workload-specific} 
\label{sec:setting}

Fast execution ensures that a placement decision arrives in time, but not that the selected page is worth moving.
Conventional policies rank pages using fixed statistics such as access counts or recency, which describe past activity rather than future migration benefit.
The same access count can indicate sustained reuse in one workload and a short-lived burst in another.
Prior work likewise finds that effective tiering thresholds and control periods vary across workloads~\cite{goodtogreat,rightchord}.
Tuning these parameters changes how aggressively a fixed rule fires, but not the statistic on which the decision is based.

We therefore explore workload-specific ranking.
This differs from warehouse-scale controllers such as TMTS~\cite{TMTS}, which seek policies that generalize across a broad workload population.
In our setting, a model can specialize to a recurring application.
Such specialization is practical because production systems repeatedly execute the same jobs and binaries~\cite{borg,googletrace,AlibabaDAG}, allowing profiling and training cost to be amortized across later runs.
Per-application learned policies also have precedent. 
Kleio trains page-placement models per application~\cite{kleio}, while learned allocators predict object behavior from application-specific allocation patterns~\cite{llama}.
\sys follows this design point, but predicts future migration benefit from current runtime observations rather than replaying a previously recorded placement.

\vspace{4pt}
\noindent\textbf{Takeaways.}
Page placement should use a workload-specific ranking policy that predicts future migration benefit from current runtime observations, rather than applying the same retrospective hotness rule to every workload.

%% file: sections/design.tex
\section{Technical Design}
\label{sec:design}

\begin{figure}[t]
  \centering
  \includegraphics[width=.85\columnwidth]{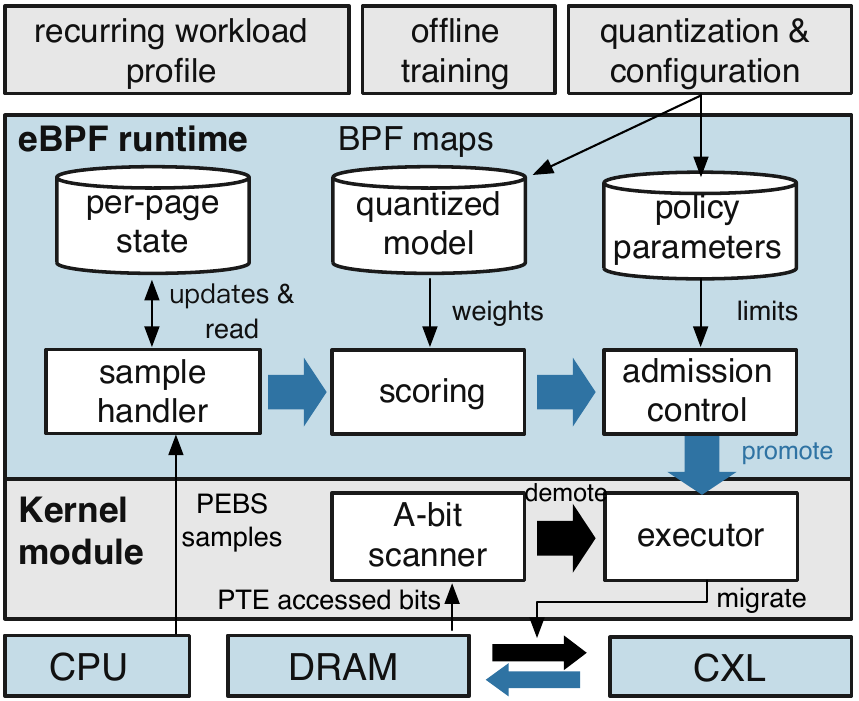}
  \caption{Design overview of \sys.}
  \label{fig:arch}
\end{figure}

\sys is an eBPF-based controller for CXL memory tiering.
Its design follows the requirements established above.
Figure~\ref{fig:arch} shows the system architecture.

First, to make timely and selective decisions, \sys observes memory accesses through PEBS and processes them in epochs of millisecond-scale.
A chain of eBPF programs maintains per-page state, constructs runtime features, and scores promotion candidates. 
High-scoring candidates do not migrate immediately.
They first pass through admission controls that bound repeated movement, per-epoch migration volume, and
fast-tier occupancy.

Second, to keep the latency-critical path in the kernel, sample processing, candidate scoring, admission, and migration execution proceed without blocking on userspace. 
Promotion intents that pass admission are
submitted to a kernel executor. 
A separate demotion path scans page-table accessed bits to identify fast-tier pages that have remained inactive and submits demotion intents to the same executor. 
The executor batches both types of intent, performs the migrations, and updates the corresponding runtime state.

Third, to customize ranking to each recurring workload, an offline training pipeline learns a compact model from a profile of the target workload. 
\sys installs the resulting quantized model, configures the eBPF programs, and publishes out-of-band parameter updates through BPF maps. 
At runtime, the model scores behavioral features of the pages observed in the current execution rather than replaying profiled addresses. 
Userspace remains outside the hot path, and kernel components continue using the most recently published configuration.
\sys manages one workload at a time, identified by its TGID. The loader attaches a PEBS event to every thread in the group. All threads then write to the same per-page state, and each page is attributed to its thread group. Concurrent management of multiple workloads is outside the scope of this paper. We discuss the extension in \S\ref{sec:discussion}.

\subsection{Event-driven telemetry and per-page state}

\sys uses PEBS samples to identify pages for potential promotion. 
Each sampled data access updates a compact state record for its virtual page. 
The record summarizes recent access counts, changes in access rate across epochs, recency, and migration history. 
These behavioral features allow the ranking policy to distinguish persistent activity from short-lived activity without depending on virtual addresses.

Algorithm~\ref{alg:promotion} in the Appendix summarizes the promotion path. 
Before a page reaches the ranking policy, the telemetry path applies inexpensive eligibility checks. 
It excludes pages already in DRAM, pages already queued for migration, and regions governed by static placement rules.
Stack pages are preferentially placed in DRAM because the active stack is typically small and frequently accessed. 
Executable regions are also handled separately because load-retired PEBS events describe data accesses rather than instruction fetches.

Eligible pages are inserted into a per-epoch candidate buffer. 
At each epoch boundary, the inference stage consumes the completed buffer while telemetry begins recording candidates for the next epoch. 
This separation keeps access collection event-driven and prevents model evaluation from blocking the telemetry path.




\subsection{Ranking promotion candidates}
\label{sec:ranking}
At each epoch boundary, \sys scores the candidate pages collected during the preceding epoch using a compact, quantized MLP. The model uses only behavioral and structural features observable at runtime, detailed in \S\ref{sec:model}. It does not use virtual addresses. The output is a scalar score that ranks a page against the other candidates in the same epoch.

The model is trained offline from a profile of the target workload. At each decision point in the profile, the pipeline records the same features available at runtime and observes the page over the following epochs. It scores the page on three properties of that window. How many accesses it receives, how many of the epochs it is touched in, and how soon the first re-access arrives. A page touched once and never again scores near zero. Raw scores are then converted to a percentile rank among the pages that were reused at all, and that rank is the value the model regresses on. The transform matters because the raw distribution is a point mass at zero with a long tail, and almost no page is reused inside the window. A model fit to raw benefit learns the true but useless fact that nearly every page is cold, and emits a near-constant score. Ranking is monotone in benefit, so it preserves the page ordering while forcing the model to spread its output across the range.

At runtime, \sys uses the model only to order candidates. The score does not trigger migration directly. It need not be a calibrated probability either. Admission consumes order, not magnitude. A rank-preserving target therefore costs nothing. A higher score gives a page higher priority, while the admission and execution policies described below determine whether and when the page is actually migrated.


\subsection{Model architecture and sizing}
\label{sec:model}  
The ranking model is a three-layer perceptron. It takes 32 inputs, passes them through hidden layers of 192 and 96 units with ReLU, and produces one scalar. Weights are signed 8-bit integers, activations unsigned, and quantization happens during training rather than after. Each layer accumulates into a 64-bit signed value, requantizes with an integer multiply and shift, then clamps through ReLU back to 8 bits. The output is read as an offset from a fixed zero point of 32, so a page with no predicted benefit lands on the zero point and higher values mean more predicted benefit. The fixed zero point also puts every score in one of 256 buckets, which is what lets the controller set each epoch's admission threshold by bucket and reduces the kernel-side test to an integer comparison.

The inputs fall into three groups. A majority describe access behavior. The current epoch's access count, a history of counts across epochs, latencies, and the differences between consecutive epochs. Two describe recency and the rest describe the virtual memory region containing the page, whether it is anonymous, whether it is shared, its three protection bits, its page type, and its size.

\subsection{Why learn instead of replay?}
\label{sec:replay}
\sys's ranking function reads how a page behaves, not which page it is. Its inputs are the access-behavior, recency, and region-attribute groups described in \S\ref{sec:model}. The virtual address is only the key of the per-page state map. It never reaches the model. Neither does the physical frame number, nor any offset derived from either. The structural inputs say what a region is, not where it sits. Address space layout randomization (ASLR) moves regions without changing their backing, permissions, or size, so these inputs do not move with it. \sys ranks whatever pages the current run produces. Allocation order, thread interleaving, and input-dependent allocation sizes change which pages hold which data. None of them change the features of a page that is being accessed at a given rate.

This is why \sys profiles a workload instead of replaying one. A recorded placement is a map from address to tier. Replaying it on a later run requires the same data to sit on the same pages, and the mechanisms above break that even when the workload, its inputs, and its binary are unchanged. The replayed profile then promotes pages that no longer hold the data that justified promoting them. A profile does hold something durable, though. \sys learns how a page's recent history predicts its future accesses. The model asks how many more accesses the page will receive, and whether that number repays the migration. That relationship belongs to the workload's computation and not to one run's address space, so it survives a change in layout. It also survives changes in phase. A recorded placement only expresses the phase in which it was recorded.

\subsection{Bounded promotion through admission control}
A model score determines a candidate's priority, not whether it moves. 
Before issuing a promotion intent, \sys applies three admission-control rules that bound movement at the page, epoch, and fast-tier levels. 
These complementary bounds limit the migration pressure caused by inaccurate or stale scores, confining their primary effect to placement quality.

First, a page that has recently migrated is excluded from reconsideration for a fixed number of epochs. 
This check occurs before model inference, avoiding both repeated scoring and rapid movement of pages whose behavior fluctuates near the admission boundary. 
Cooldown therefore bounds how frequently any individual page can move between tiers.

Second, among eligible candidates, \sys admits only the candidates whose scores exceed a threshold derived from the recent score distribution. 
The userspace controller computes this threshold from the score histogram of the completed epoch and publishes it for the next epoch. 
This histogram-based implementation approximates top-\(K\) selection without sorting candidates in the kernel. 
It bounds the number of promotion intents while favoring the best opportunities available in the current workload phase.

Third, before creating an intent, \sys checks the current fast-tier occupancy against its configured target. 
Promotions stop when the target is reached and resume after demotion creates capacity. 
This prevents a high-scoring candidate stream from overcommitting the fast tier.

\subsection{Adaptive sampling decision cadence}

\sys uses a two-state controller to reduce decision overhead after placement converges. 
At startup, the controller enters \textsc{Active} mode, which uses a short epoch and a low PEBS sampling period to collect samples and evaluate candidates frequently. 
At each epoch boundary, the controller observes the candidate rate, admitted promotion rate, and number of cross-tier page movements.

When all three signals remain below their convergence thresholds for \(N_{\mathrm{stable}}\) consecutive epochs, the controller transitions to \textsc{Cruise} mode. 
In this mode, it increases both the PEBS sampling period and the epoch duration, reducing telemetry and inference work when few beneficial migration opportunities remain.
If the candidate rate or cross-tier movement rises above its reactivation threshold, the controller returns to \textsc{Active} mode. 
Requiring sustained stability before entering \textsc{Cruise} prevents short-term fluctuations from repeatedly switching the controller between modes.

The controller changes how frequently \sys observes and scores pages, but not the ranking model or admission policy. 
It therefore reduces steady-state overhead while retaining the ability to detect renewed placement activity.




\subsection{Conservative A-bit demotion}
\label{sec:demotion}

\sys handles promotion and demotion asymmetrically. Promotion must respond quickly to newly active CXL pages, for which PEBS provides precise positive access samples. 
Demotion should be more conservative.
Moving a page out of DRAM prematurely causes subsequent accesses to pay the CXL latency penalty. 
\sys therefore bases demotion on absence of access, using the page-table Accessed bit as a low-cost signal.

Algorithm~\ref{alg:demotion} in the Appendix summarizes the demotion path. 
Once the scanner is active, it traverses the page tables in chunks of 50{,}000 entries, about once a second, and examines DRAM-resident pages. A set A-bit means the page was touched since the last visit, so the scanner clears the bit and leaves the page in DRAM. A page whose bit is still clear has gone a full scan period without a single access, and is emitted as a demotion intent to the shared migration executor. 
Clearing the bit on every visit is what makes the test conservative.
A page must be untouched across the whole interval, not merely unsampled.

The scanner resumes each traversal from its previous stopping point rather than walking the complete address space at once. 
This bounds the work performed in each scan while eventually revisiting all DRAM-resident pages.





\subsection{Batched migration execution}
The migration executor separates latency-critical placement decisions from the substantially more expensive page migration path. Promotion and demotion policies submit intents to a bounded lock-free queue, so telemetry and ranking continue while earlier decisions are executed. The executor drains the queue in batches, bounded by a per-cycle migration budget so a burst of intents cannot monopolize the worker thread. Each batch is sorted by virtual address and adjacent pages are merged into ranges, so a single migration call and shootdown can cover many pages. A page's tier state is updated only after its migration succeeds.

The executor also enforces fast-tier capacity. It tracks the net number of pages it has placed in the fast tier and promotes only while that count is below the configured target; demotions continue regardless and return capacity. The demotion scanner is driven by the same counter and starts at 70\% of the target, so space is freed before the cap is reached. The counter advances only when a migration completes and the promotion test is sampled once per drain cycle, so occupancy can exceed the target by at most one cycle's migration budget. When the queue is empty the executor backs off its polling, lowering steady-state CPU overhead.





%% file: sections/impl.tex
\section{Implementation}
\label{sec:impl}
\sys consists of approximately 800 lines of eBPF code for telemetry and inference, a 900-line kernel module for demotion and migration, and a 1.5K-line userspace loader for setup and control. 
The components exchange configuration, model parameters, page state, candidates, and migration intents through pinned BPF maps. 
The implementation is publicly available under GPLv2 at https://github.com/a85tract/xtier.

\vspace{4pt}
\noindent \textbf{Verifier-compatible inference.} 
The loader attaches PEBS perf events to every monitored thread. Each event invokes a chain of eBPF programs that updates page state and evaluates a 32-input MLP with hidden layers of 192 and 96 units and one scalar output. The complete forward pass is too large for a single verified BPF program, so \sys partitions it into tail-called stages and stores intermediate activations in a per-CPU scratch map. The quantized weights occupy 114 KB of map state.
Per-page state resides in a bounded LRU hash map keyed by page-frame number. 
Candidate buffers and score histograms are double-buffered.
Telemetry populates one buffer while inference or threshold computation consumes the other, and the buffers swap at epoch boundaries. 
This allows PEBS processing to continue without waiting for the preceding epoch's inference and control work.



\vspace{4pt}
\noindent \textbf{Kernel migration support.}
Page migration requires MM operations that are unavailable to eBPF and not exported to loadable modules at the required granularity. We add a single wrapper, \texttt{xtier\_migrate\_range}, to \texttt{mm/mempolicy.c} in Linux 6.11. Given an \texttt{mm\_struct}, virtual-address range, and target NUMA node, the wrapper invokes the kernel's existing migration machinery. This wrapper is the only modification to the base kernel.

Intents reach the module through a lock-free ring of 32{,}768 entries, appended to with a compare-and-swap on the tail. A page already enqueued this epoch is dropped, so one hot page cannot fill the ring. When the ring is full the producer drops the intent and records it. Blocking would stall a PEBS handler. A dropped intent costs little, because the page is scored again next epoch if it is still active. The module drains at most 64 intents per cycle, sorts them by address, and merges contiguous runs so one migration call and one shootdown cover each run. Merging stops at the first gap, so a range spans at most the per-cycle budget. Tier state is updated when the migration call returns, at range granularity, and a failed range leaves it untouched.


\vspace{4pt}
\noindent \textbf{Userspace control and training.} 
At startup, the loader installs the quantized model, populates region metadata from \texttt{/proc/\allowbreak <pid>/maps}, configures PEBS, and starts the adaptive controller. During execution, it derives the next epoch's admission threshold from the completed score histogram and publishes the result through a BPF map. These operations are outside the per-decision path. If the loader is delayed, kernel components continue using the most recently published configuration.

The offline pipeline profiles the workload for 15 minutes, labelling candidates as described in \S\ref{sec:ranking}. Training on that trace takes one to two minutes on a single CPU. We train with Huber loss and AdamW, a learning rate of 0.005 under cosine annealing, and batches of 1024. Redeploying a retrained model replaces the weight header and rebuilds the loader. The eBPF object is unchanged, because the loader pushes weights into BPF maps at startup.

%% file: sections/eval.tex
\section{Evaluation}
\label{sec:evaluation}

\begin{figure*}[th]
  \centering
  \includegraphics[width=\textwidth]{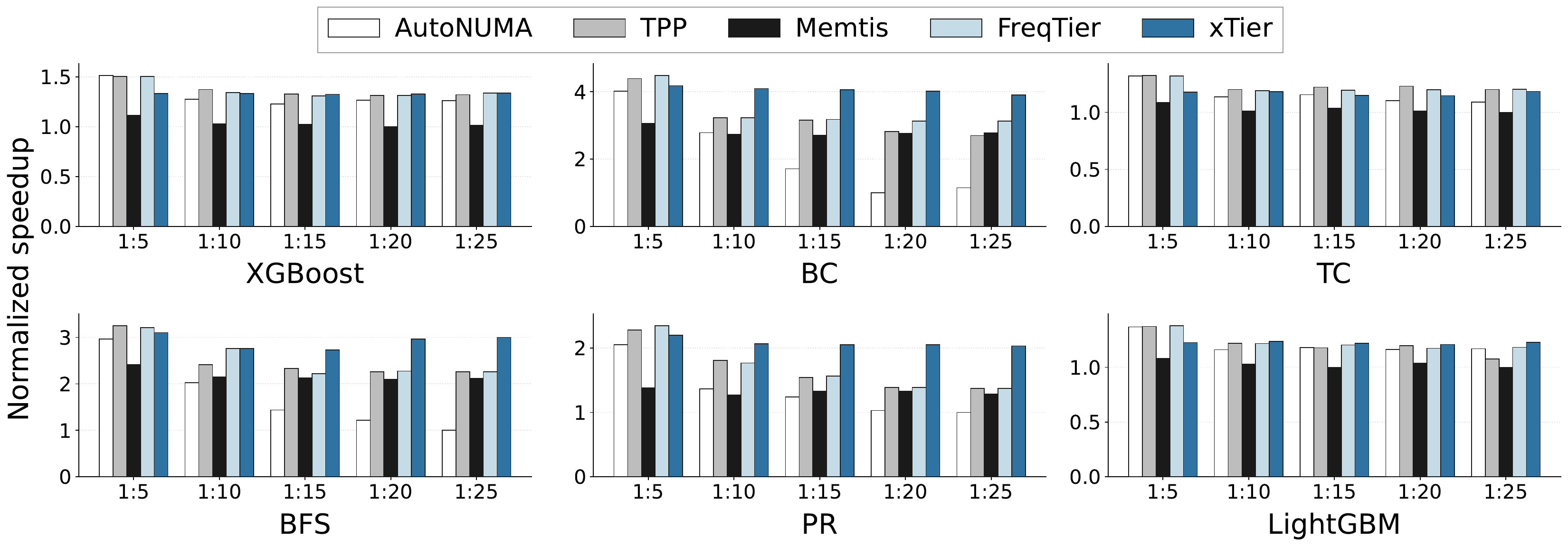}
  \caption{Normalized speedup across workloads and DRAM:CXL ratios. For each workload, the speedup is normalized to the slowest system--ratio configuration. Higher is better.}
  \label{fig:speedup-grid}
\end{figure*}

We evaluate \sys to answer four questions:
\textbf{RQ1}: How does \sys compare to existing tiered-memory systems?
\textbf{RQ2}: Where do \sys's gains come from?
\textbf{RQ3}: Does \sys's learned policy generalize and adapt over time?
\textbf{RQ4}: What are the costs of deploying \sys?

\subsection{Methodology}
\label{sec:methodology}
\noindent \textbf{Hardware platform and memory tiers.}
We run all experiments on CloudLab c220g2 machines~\cite{Cloudlab}, each with two 10-core Intel Xeon E5-2660 v3 sockets and roughly 80\,GB of DDR4 per NUMA node.
  Following prior work~\cite{Memtis,TPP,HeMem,FreqTier}, we pin workloads to socket~0 and treat its memory as DRAM and socket~1 memory as the slow tier.
  The \texttt{memmap} kernel parameter caps DRAM to impose each capacity ratio.
  Measured idle load-to-use latency is 114\,ns local and 143\,ns remote, a $1.25\times$ ratio, against roughly $2\times$ at idle and $2.9\times$ under load on CXL 1.1 hardware~\cite{Pond,FreqTier}.
  We therefore evaluate \sys under a smaller tier-latency gap than real devices impose.
  Unless otherwise noted, all experiments use Linux 6.11 with identical software, workloads, and inputs, varying only the placement policy.

\vspace{4pt}
\noindent \textbf{Workloads.}
We evaluate two classes of  memory-intensive workloads that are commonly used in tiering studies~\cite{TPP,HeMem,FreqTier,Memtis}.
The first consists of gradient-boosted tree training, represented by XGBoost~\cite{xgboost} and LightGBM~\cite{lightgbm}. 
The second consists of graph analytics: Triangle Counting (TC), Betweenness Centrality (BC), Breadth-First Search (BFS), and PageRank (PR) from GAPBS~\cite{gapbs} on a Kronecker power-law graph.
Unless otherwise stated, each workload's resident set is at least $5\times$ the largest evaluated DRAM capacity.



\vspace{4pt}
\noindent \textbf{Baselines.} 
We compare \sys with AutoNUMA~\cite{AutoNUMA}, TPP~\cite{TPP}, Memtis~\cite{Memtis}, and FreqTier~\cite{FreqTier}, using their recommended configurations unless otherwise noted. 
These systems represent upstream Linux mechanisms and recent policies for CXL-class tiered memory. We omit systems targeting fundamentally different memory technologies, such as those designed for NVM or network-attached far memory. These are calibrated to slow-tier latencies several times higher than CXL, so a head-to-head comparison would conflate effects we cannot separately attribute.

\vspace{4pt}
\noindent \textbf{Metrics.} 
We report application runtime, fast-tier occupancy, migration volume, and \sys's decision-path overhead. Runtime uses training time for XGBoost and LightGBM and kernel execution time for GAPBS. Fast-tier occupancy is sampled from \texttt{numa\_maps}, while migration volume and decision latency are collected via kernel and eBPF instrumentation.





\subsection{RQ1: How does \sys compare to existing tiered-memory systems?}
\label{sec:rq1}

Figure~\ref{fig:speedup-grid} compares the end-to-end performance of AutoNUMA, TPP, Memtis, FreqTier, and \sys across the six workloads and five DRAM ratios. 
Overall, \sys delivers the best performance in most configurations and remains close to the best-performing baseline when it does not. 
At larger DRAM fractions, the differences among systems are generally smaller, and several baselines occasionally match or outperform \sys.

The advantage of \sys becomes more pronounced as the DRAM budget shrinks. 
Under tighter fast-tier capacity, \sys is consistently competitive and often outperforms the baselines, particularly for BC, BFS, PR, and LightGBM. 
These configurations leave less room for poor placement decisions and migration churn, making effective use of the limited DRAM capacity increasingly important.

\begin{figure*}[t]
  \centering
  \includegraphics[width=\textwidth]{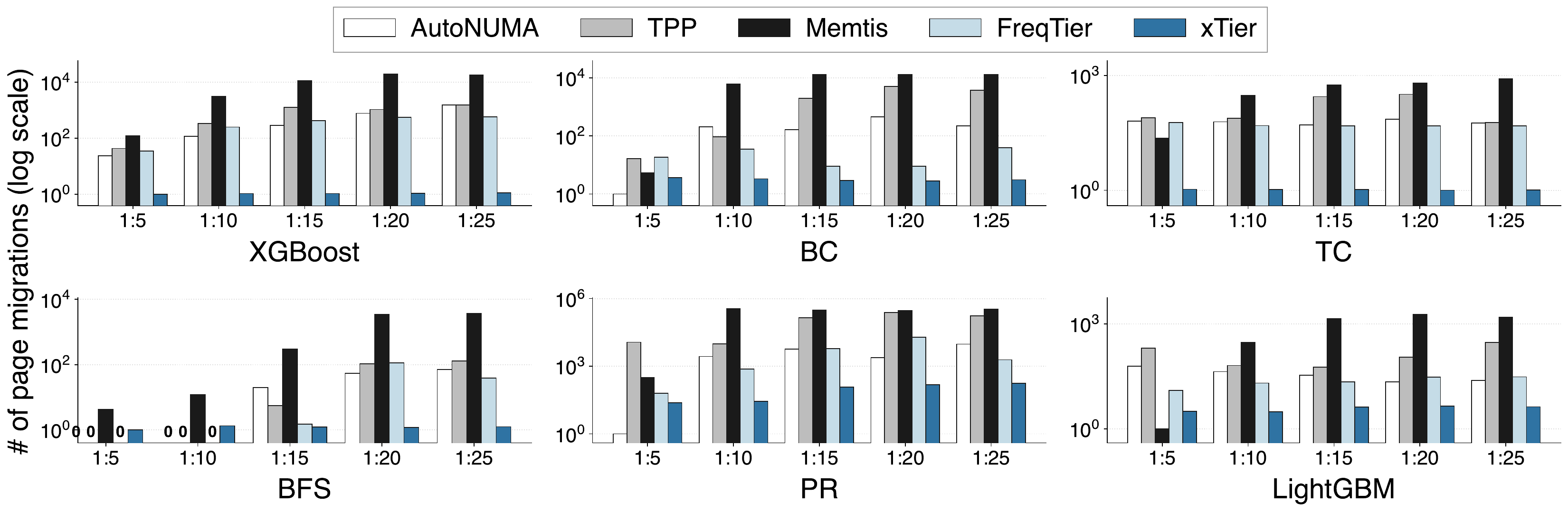}
  \caption{Page migrations across workloads and DRAM:CXL ratios. Lower is better. The y-axis is logarithmic. A ``0'' label indicates that no migrations were observed for that system--ratio configuration.}
  \label{fig:migrations-grid}
\end{figure*}

Overall, RQ1 shows that \sys maintains strong end-to-end performance across both workload classes, with its largest advantages appearing when fast-tier capacity is constrained. End-to-end performance alone, however, does not reveal whether these gains come from better page placement, less migration work, or both. RQ2 separates these effects.

\subsection{RQ2: Where do \sys's gains come from?}
\label{sec:rq2}

RQ1 establishes that \sys achieves strong end-to-end performance, but does not identify which parts of the design are responsible. 
The promotion path has two distinct components.
The learned ranker determines which candidate pages are most valuable to promote, while the filter stack bounds which samples require ranking and which promotions are allowed to proceed. 
We isolate their roles through ranker ablation and filter-removal experiments.

\vspace{4pt}
\noindent \textbf{Ablation.}
We first replace the learned ranker with a frequency based score while leaving the rest of the promotion pipeline unchanged. 
We then repeat both ranker configurations with the filter stack removed. 
As a reference point, we also measure a static configuration that leaves the workload on CXL and performs no tiering.

With the filters enabled, both rankers outperform static CXL placement. 
This shows that the surrounding promotion machinery remains useful even with a simple frequency score. 
The learned ranker performs better still, and its advantage over frequency persists across all DRAM ratios.

The behavior changes sharply when the filters are removed. 
Neither ranker completes within our fixed cutoff. 
Telemetry collection itself costs only 345~ns per PEBS sample, whereas scoring one candidate with the deployed MLP costs 34~$\mu$s, nearly two orders of magnitude more.
Without the pre-ranking checks, every PEBS sample incurs scoring cost, including samples for pages already in DRAM or recently considered for migration. 
Removing cooldown and admission controls also allows migration requests to accumulate faster than they can be drained, causing repeated movement between tiers and eventually preventing the workload from making useful progress.

These results separate the roles of the two components. 
The filters make ranking operationally affordable by restricting the 34~$\mu$s inference to samples whose placement is still actionable.
For candidates that are ultimately admitted, page migration adds another 19~$\mu$s, bringing scoring and movement to roughly 53~$\mu$s per admitted promotion. 
Avoiding unnecessary inference is therefore important even before migration cost is considered. 
Once the filters have bounded the decision space, performance depends on how well the remaining candidates are ranked.

Within that candidate set, the learned model consistently outperforms frequency. 
Grouped feature-importance analysis provides some insight into this difference. 
The two most influential feature groups are the per-epoch access deltas and the multi-epoch access-count history, both of which capture how access intensity evolves over time rather than only its accumulated magnitude. 
This is consistent with the prediction task, as changes across recent epochs can distinguish pages whose demand is rising from pages whose past frequency is high but whose demand is already fading. 
A frequency score observes mainly what a page has already accumulated, whereas the learned ranker can incorporate the direction and recent history of that demand.

\vspace{4pt}
\noindent \textbf{Quantization.} 
Because the deployed model is quantized (\S\ref{sec:impl}), we also verify that INT8 conversion preserves the decisions made by the ranker.
Across held-out pages, quantization reduces F1 by at most 1.5 percentage points. 
More importantly for \sys, the Spearman correlation~\cite{spearman} between float and INT8 scores remains above 0.98 in every workload and DRAM configuration. 
Since \sys uses these scores primarily to order promotion candidates, this high rank correlation indicates that quantization preserves the ordering on which admission decisions depend.


\vspace{4pt}
\noindent \textbf{Migration volume.}
  Figure~\ref{fig:migrations-grid} shows the placement work underlying the end-to-end results from RQ1.
  Across workloads and DRAM:CXL ratios, \sys generally performs the fewest migrations, and the gap widens as the DRAM fraction shrinks.
  While several baselines increase migration activity by orders of magnitude, \sys keeps page movement comparatively low.
  A zero migration count does not imply that the entire resident set fits in DRAM.
  For some workload--ratio combinations, the actively accessed footprint can be served without additional migrations.

  Migration is not free.
  Each move costs CPU time, memory bandwidth, and address-translation maintenance, so at high rates it becomes a substantial part of tier-management cost.
  The results thus explain the performance observed in RQ1.
  The filter stack bounds how much decision and migration work happens, while the learned ranker improves which pages receive that budget.
  The ablation shows the same split.
  Removing the filters produces unsustainable migration pressure, while the frequency ranker with filters intact stays operational but underperforms learned ranking.

  Fewer migrations are not the objective, and migration count does not translate linearly into speedup.
  The important result is that \sys matches or beats end-to-end performance while moving substantially fewer pages.
  Its gain comes from making migration more selective, not from reducing activity for its own sake.

\begin{figure}[t]
  \centering
  \includegraphics[width=\columnwidth]{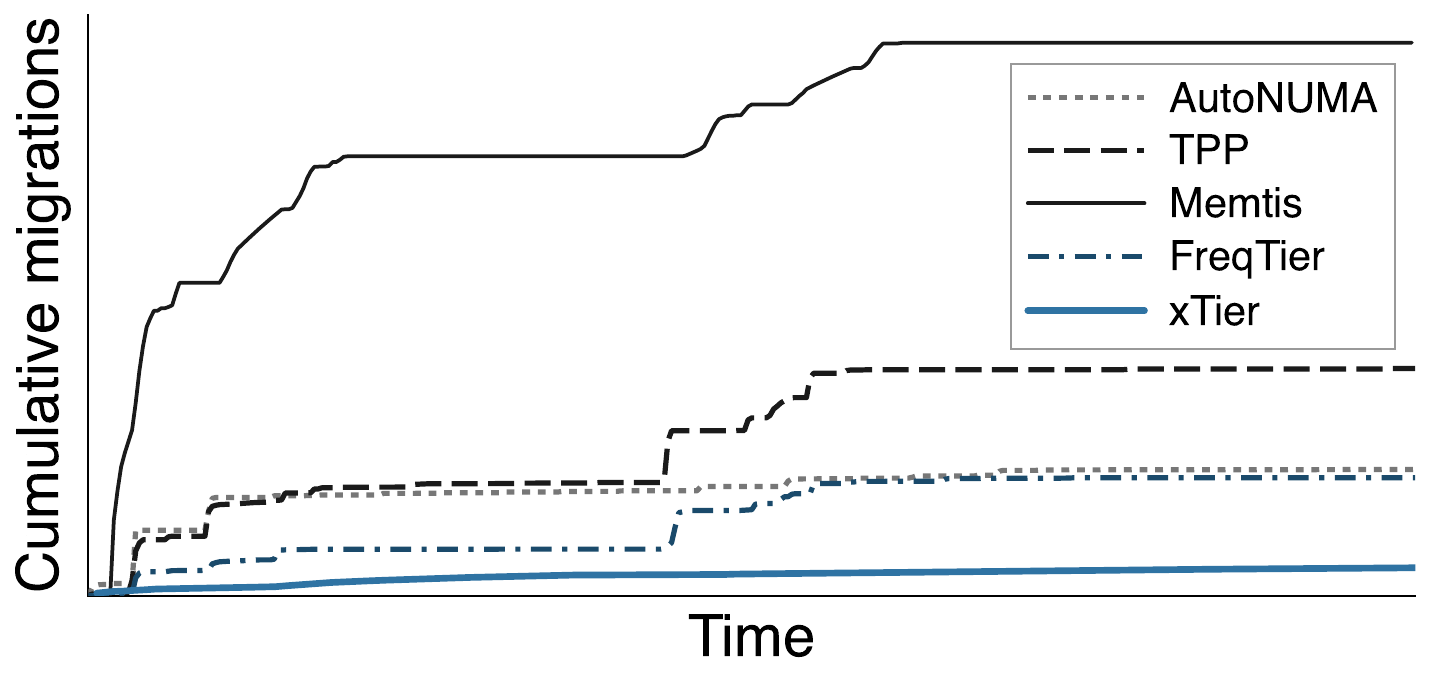}
  \caption{Cumulative page migrations over time for XGBoost. \sys performs migration early in execution and then approaches a low steady-state rate, while several baselines exhibit additional migration bursts later in the run.}
  \label{fig:migrations-time}
\end{figure}




\vspace{4pt}
\noindent \textbf{Ranking quality.} 
We also evaluate whether the learned model produces a useful ordering independently of the rest of the promotion policy.
We use a temporal split, training on the first 70\% of execution epochs and testing on the final 30\%.
This prevents future execution behavior from leaking into training, as could occur with a random split over samples from the same run.
We report AUC because it evaluates ranking without choosing an admission threshold. It measures how often a page that would benefit from promotion outscores one that would not.

The deployed INT8 ranker achieves AUCs of 0.767 on LightGBM, 0.791 on XGBoost, and 0.848 on PageRank, compared with 0.5 for an uninformative ranker.
The model provides useful discrimination across all three workloads, although ranking quality alone does not determine end-to-end benefit.
The value of precise ordering also depends on how concentrated promotion benefit is.
For LightGBM, a small fraction of pages accounts for most of the available benefit, so identifying the highest-value pages is important.
For PageRank, benefit is distributed much more broadly.
Roughly one third of the pages would be required to capture 90\% of it.
In this regime, many candidates have similar value, so controlling admission volume matters more than distinguishing finely among them.
This explains why ranking quality and end-to-end speedup do not necessarily move together.
When benefit is concentrated, ranking matters most.
When benefit is diffuse, bounded admission becomes more important.

\begin{figure}[t]
\centering 
\includegraphics[width=\columnwidth]{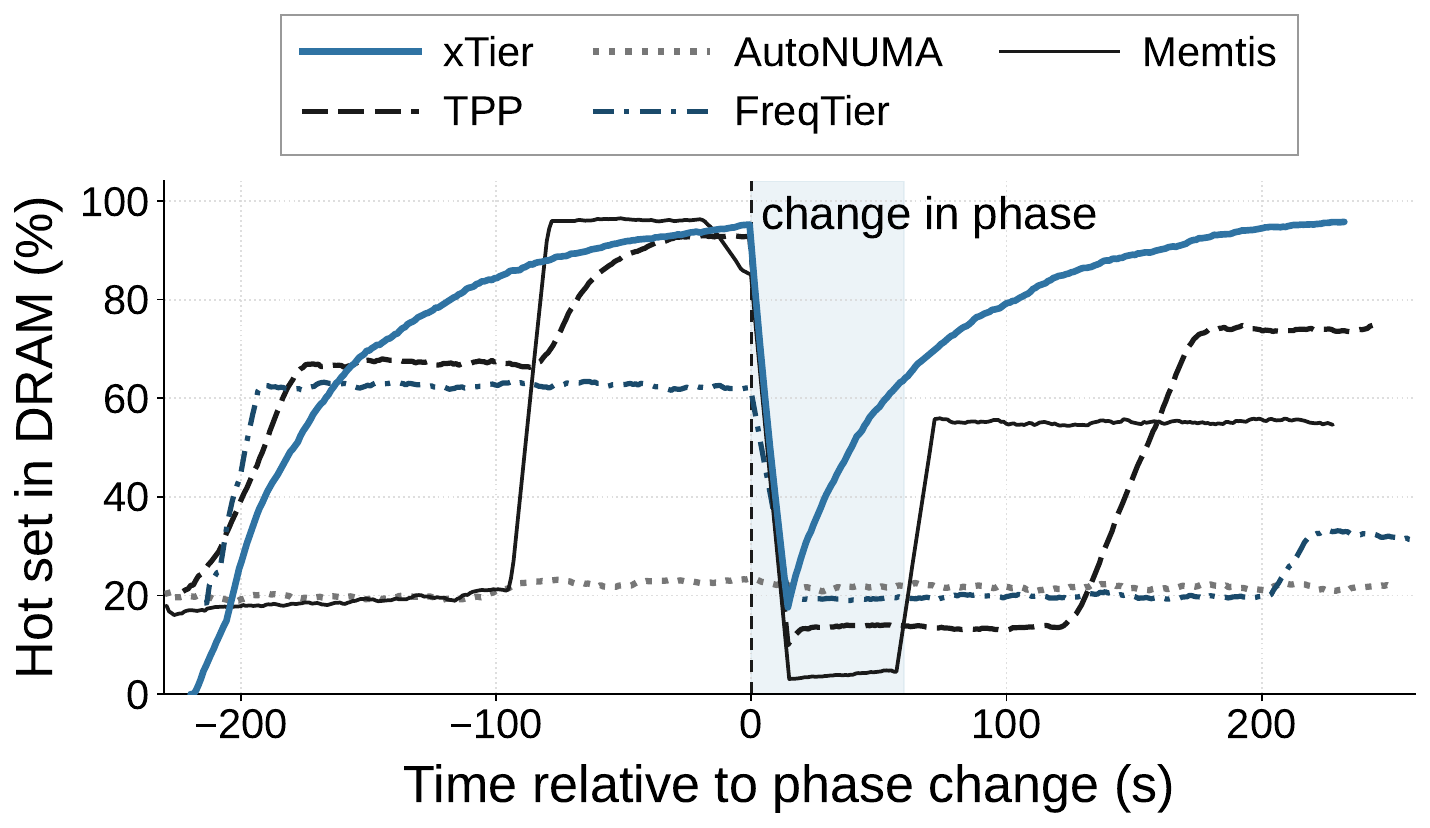}
\caption{DRAM coverage of the currently hot segment around a forced phase change. The shaded region marks the first 60s after the change.}
\label{fig:phase} 
\end{figure} 

\subsection{RQ3: Does \sys's learned policy generalize and adapt over time?}
\label{sec:rq3}

\vspace{4pt}
\noindent \textbf{Convergence over time.}
RQ2 showed that \sys performs fewer migrations in aggregate.
Figure~\ref{fig:migrations-time} shows how this difference develops over time for XGBoost.
Memtis and TPP exhibit step-like cumulative curves, with later bursts of migration after their initial placement activity.
AutoNUMA and FreqTier are less bursty, but also continue to incur additional migration work after the initial phase.

\sys behaves differently.
Most of its migration activity occurs during initial placement discovery, after which the slope of its cumulative migration curve falls sharply.
Samples continue to arrive, but pages already handled, low-value candidates, and candidates rejected by admission control do not translate into page movement.
As a result, migration activity approaches a low steady-state rate once the useful placement opportunities in the current phase are exhausted.

\vspace{4pt}
\noindent \textbf{Phase response.} 
Convergence alone, however, is insufficient.
A tiering policy must also leave a converged placement when the workload changes.
We therefore use a controlled microbenchmark to test this behavior. We use \texttt{zipf\_chase}, which relocates its hot set to a disjoint region partway through execution.
Figure~\ref{fig:phase} reports the fraction of the currently hot region resident in DRAM around this phase change.
Because the newly hot pages initially reside on CXL, coverage drops to nearly zero for every system at the transition.
\sys begins recovering within a few seconds and increases steadily to 94\% coverage.
In contrast, the baselines exhibit long plateaus before substantial replacement occurs.
Memtis shows little improvement for roughly 60~s before reaching 55\%, TPP for 131~s before 74\%, and FreqTier for 216~s before  30\%.

The plateaus indicate that pages useful to the previous phase remain in DRAM for a substantial period before enough capacity is reclaimed for the new hot set.
\sys replaces this stale placement more continuously.
Ranking and admission repeat every epoch, and demotion frees pages that have gone cold. New-phase samples can therefore win promotions as soon as they are competitive.
The result is a gradual recovery rather than a delayed burst of replacement.
Thus, convergence does not make \sys static.
Migration activity falls during stable phases, rises again when the working set changes, and falls once the new placement converges.

\vspace{4pt}
\noindent \textbf{Generalization across inputs.}
The model is trained offline, so its usefulness depends on whether a profile collected from one execution transfers to later runs of the same workload.
We pick two different inputs per workload and train a model on each. Each input is then scored by both models.
We compare the two models on the same target input so that differences in input difficulty do not affect the comparison.
The paired inputs also use the same memory footprint to keep sampling density comparable.

At the deployed admission rate, swapping the model changes precision by only 0.4\% for Light\-GBM, 0.1\% for PageRank, and 0.3\% for XGBoost.
These differences are negligible, indicating that the learned ranking transfers well across the tested inputs of the same workload.
This supports the intended deployment model in which profiling cost is amortized across recurring runs rather than repeated for every input.

\subsection{RQ4: What are the costs of deploying \sys?}
\label{sec:rq4}
The previous sections show that \sys achieves strong end-to-end performance while making page migration more selective.
We now quantify the cost of \sys in terms of memory footprint, CPU overhead, and offline training cost.





\vspace{4pt}
\noindent \textbf{Memory footprint.}
\sys uses approximately 218\,MB of resident memory in steady state, almost all of which is BPF map state.
The dominant component is the per-page state map, which stores recent access history, recency, and tier metadata.
By comparison, the deployed INT8 MLP occupies only 114\,KB, and its size is fixed by the model architecture.

The 218\,MB footprint is an additional cost relative to policies such as AutoNUMA and TPP, which largely reuse existing kernel NUMA and reclaim metadata.
On our platform, however, it corresponds to only 0.27\% of one 80\,GB NUMA node and 0.14\% of total system memory.
Thus, the metadata cost is small relative to the memory capacity being managed.

\vspace{4pt}
\noindent \textbf{CPU overhead.}
The dominant runtime costs come from PEBS telemetry and candidate scoring.
As discussed in RQ2, processing a PEBS sample and updating its page state costs 345\,ns, whereas an MLP inference costs 34\,\textmu s.
These operations occur at very different frequencies.
Every sample incurs telemetry cost, while only candidates that pass the pre-ranking filters are scored.
Across our workloads, fewer than one fifth of samples reach inference, and fewer than one percent produce an admitted promotion.
After placement stabilizes, the fraction reaching inference can fall below one percent because most samples are rejected before scoring.

A sweep over PEBS sampling frequency shows the resulting tradeoff.
At low sampling rates, CPU consumption remains near 2\% of one core, where fixed per-epoch work dominates.
As the sampling rate increases, CPU use grows approximately linearly because more samples enter the BPF path and more candidates reach inference.
At the highest rates tested, utilization exceeds 25\% of one core, confirming that unrestricted high-rate sampling would be too expensive.

Normal operation avoids this regime.
Once placement stabilizes, the adaptive controller reduces the sampling rate and lengthens the epoch interval.
Across our workloads, steady-state CPU utilization remains below 4\% of one core, with higher utilization appearing primarily during startup and phase transitions.
The migration executor contributes less than 0.02\% of one core in all tested configurations.
Thus, \sys controls runtime overhead primarily by limiting how often expensive decisions are invoked rather than by making each decision itself inexpensive.

\vspace{4pt}
\noindent \textbf{Training cost.}
Model training occurs offline and is paid once for each workload profile.
\sys collects a profiling trace, derives supervision from subsequent accesses, trains the MLP, quantizes it to INT8, and exports the weights consumed by the runtime loader.
Across the workloads evaluated, this process requires 1--2 minutes of model training on a single CPU and produces a 114\,KB model.
The training cost is therefore outside the runtime critical path and can be amortized across subsequent executions of the same recurring workload.

%% file: sections/discussion.tex
\section{Discussion and Future Work}
\label{sec:discussion}

\noindent \textbf{Online learning.} \sys currently trains a workload-specific ranking model offline and holds it fixed during execution. 
Updating the model directly in eBPF would conflict with verifier constraints, integer-only computation, and the latency requirements of the sampling path. 
A practical extension would retain inference in the kernel while performing training and lightweight adaptation asynchronously in userspace, with updated weights validated before deployment.


\vspace{4pt}
\noindent \textbf{Multiple workloads.}
\sys currently assumes a single managed workload.
The mechanism extends naturally to multiple workloads.
The harder problem is arbitration over a shared fast tier.
Sound decisions can conflict when workloads compete for capacity. Scores from separate models are also not comparable, because each is meaningful only against its own workload's distribution.
A multi-workload design therefore requires an explicit cross-workload policy that incorporates fairness, priority, or QoS objectives depending on the deployment.
We leave the design and evaluation of these policies under contention to future work.

\vspace{4pt}
\noindent \textbf{Generalization.}
Offline specialization suits managed deployments where memory-intensive workloads are versioned, scheduled and repeatedly executed. Its benefit may diminish for one-off workloads, or when access behavior changes between profiling and deployment. Admission control still bounds migration volume, but inaccurate rankings reduce placement quality. The shift lies in how features relate to future reuse, not in the features themselves. A detector watching the input distribution would miss it. Future work could instead compare predicted against observed reuse for promoted pages, and retrain from that.


\vspace{4pt}
\noindent \textbf{Hardware dependence.}
Our prototype uses Intel PEBS as precise positive evidence of memory accesses and page-table Accessed bits as inexpensive evidence of inactivity. 
Porting \sys to other architectures requires mapping these signals to the available sampling and page-reference mechanisms. 
The design itself is not specific to PEBS.
Promotion requires selective evidence of active pages, whereas demotion benefits from a cheaper, conservative signal of inactivity.


\vspace{4pt}
\noindent \textbf{CXL emulation.}
Following prior tiered-memory studies, we use remote NUMA memory to compare all policies under identical hardware and software conditions. 
Our measured $1.25\times$ tier-latency ratio is smaller than the approximately $2\times$ idle and $2.9\times$ loaded ratios reported for CXL 1.1 systems. 
However, NUMA emulation does not reproduce device-side contention, topology, or bandwidth behavior of physical CXL memory. 
Evaluation on multiple generations of CXL hardware is therefore needed to establish how these effects influence \sys's absolute and relative performance.


%% file: sections/related.tex
\section{Related Work}
\sys builds on prior work in tiered-memory placement, in-kernel machine learning, and BPF kernel customization.

\vspace{4pt}
\noindent \textbf{Tiered-memory placement.} 
AutoNUMA~\cite{AutoNUMA} and kernel resident tiering systems such as TPP~\cite{TPP}, Memtis~\cite{Memtis}, and FreqTier~\cite{FreqTier} promote and demote pages using hand-designed signals, including NUMA hint faults, access counts, and frequency thresholds. 
In contrast, \sys learns a workload-specific ranking over access history, recency, and structural context, while adapting both sampling and decision cadence as placement converges. 
Other systems put policy in userspace: HeMem~\cite{HeMem} uses PEBS samples, while Kleio~\cite{kleio}, ArtMem~\cite{artmem}, and FarSight~\cite{farsight} apply learned policies. 
Although userspace supports flexible policy logic, it introduces kernel-boundary and scheduling overheads that \sys avoids by placing inference on the in-kernel decision path.


\vspace{4pt}
\noindent \textbf{Machine learning inside the kernel.} 
Prior systems have brought machine learning close to kernel execution. KML~\cite{KML} uses a custom in-kernel framework for system-wide storage tuning, while LAKE~\cite{lake} remotely invokes GPU accelerators from kernel space. O2C~\cite{O2C} embeds decision-tree inference in eBPF, and related work applies lightweight models to kernel packet processing~\cite{KernelPacketML,kernelnn}. 
\sys instead executes a quantized neural network entirely within eBPF for page placement. Learning has also improved cache prediction, prefetching, and replacement outside the kernel~\cite{map,Pythia,tmap,longuvm,Glider,lecar}.
\sys applies the same broader principle to online tiered-memory ranking.


\vspace{4pt}
\noindent \textbf{BPF kernel customization.} 
FetchBPF~\cite{fetchbpf}, cache\_ext~\cite{cacheext}, PageFlex~\cite{pageflex} and eBPF-mm~\cite{ebpfmm} expose extensible policies for prefetching, page cache eviction, and page-size promotion, respectively. 
Their policies are hand-authored or driven by userspace profiles. 
\sys extends the deployment model to learned tiered-memory placement, demonstrating that neural inference and its placement decisions can remain on the in-kernel path.

\vspace{4pt}
 \noindent \textbf{Complementary axes.} 
Recent systems improve tiered memory along dimensions complementary to \sys.
Alto~\cite{alto} and PACT~\cite{pact} replace access frequency with richer latency- or stall-based signals; such signals are orthogonal to our learned ranking and could be incorporated as model features.
NeoMem~\cite{neomem} offloads profiling to a CXL device-side controller, while ARMS~\cite{arms}, FlexMem~\cite{flexmem}, and MTM~\cite{mtm} adapt profiling or migration cadence at runtime.
Recent studies likewise show that effective thresholds and control periods vary across workloads~\cite{goodtogreat,rightchord}.
These approaches improve how a fixed policy is observed or tuned, whereas \sys learns the page-ranking policy itself.
TierBPF~\cite{tierbpf} also executes tiering policy through eBPF, but exposes hooks for hand-written admission rules rather than learned per-page ranking and in-kernel inference.
Outside memory tiering, 3L-Cache~\cite{3lcache} similarly studies how to make a learned online policy inexpensive enough for the runtime hot path, in the context of software cache eviction.

%% file: sections/conclusion.tex
\section{Conclusion}
\label{sec:conclusion}

CXL memory expansion turns page placement into a first-order operating-system problem. Existing systems either put flexible policy in userspace, where scheduling and boundary crossings delay decisions, or use fixed kernel heuristics that cannot adapt to a particular workload. This paper presented \sys, a kernel-resident learned tiering controller that uses eBPF and PEBS to score promotion candidates inside the kernel.

\sys treats DRAM as an admission-controlled tier. A page moves only when a compact learned model and runtime gates agree that the move will repay its cost. Across six workloads and five DRAM:CXL ratios, \sys matches or beats existing policies while moving far fewer pages. Effective CXL tiering is not a matter of migrating more aggressively. It is a matter of deciding well and bounding what those decisions cost.

%% file: sections/appendix.tex
\section{Algorithms}
\subsection{Promotion}

\begin{algorithm}[tb]
\small
\caption{Hot page promotion}
\label{alg:promotion}
\DontPrintSemicolon
\LinesNumbered

\KwIn{PEBS samples}
\KwOut{Batched promotion intents}
\ForEach{PEBS sample}{
    $p \leftarrow$ sampled page\;
    \If{$p$ is stack or executable}{
        Skip $p$ \tcp*[f]{static placement}
    }
    Update page state for $p$\;
    \If{$p$ is already on DRAM}{
        Skip $p$ \tcp*[f]{no promotion needed}
    }
    
    \If{$p$ was promoted recently}{
        Skip $p$ \tcp*[f]{avoid thrashing}
    }
    $s_p \leftarrow$ ML score for $p$\;
    \If{$s_p > \tau_K$}{
        Add $p$ to promotion batch\;
    }
}

\BlankLine
Submit promotion batch to migration executor\;
\end{algorithm}

\subsection{Demotion}

\begin{algorithm}[tb]
\small
\caption{Cold page demotion}
\label{alg:demotion}
\DontPrintSemicolon
\LinesNumbered
\KwIn{Page-table scan}
\KwOut{Batched demotion intents}
\If{DRAM occupancy $<70\%$ capacity}{
    Stop scan \tcp*[f]{fast tier has free space}
}
\ForEach{DRAM page $p$ in scan chunk}{
    \If{Accessed bit of $p$ is set}{
        Clear Accessed bit\;
        Skip $p$ \tcp*[f]{recently used}
    }
    \If{Accessed bit of $p$ was clear in previous scan}{
        Add $p$ to demotion batch\;
    }
}
\BlankLine
Submit demotion batch to migration executor\;
\end{algorithm}